\documentclass[aps,prl,twocolumn,groupedaddress]{revtex4}
\usepackage{graphicx}

\newcommand{\bwt}{\begin{widetext}}
\newcommand{\ewt}{\end{widetext}}
\newcommand{\bea}{\begin{eqnarray}}
\newcommand{\enea}{\end{eqnarray}}
\newcommand{\beq}{\begin{equation}}
\newcommand{\eneq}{\end{equation}}

\begin{document}
\title{Observation  of mesoscopic conductance fluctuations in YBaCuO grain boundary Josephson Junctions}
\author{A. Tagliacozzo$^{1}$, D. Born$^{1}$, D. Stornaiuolo$^{1}$, E. Gambale$^{2}$, D. Dalena$^{2}$, \\
F. Lombardi$^{3}$, A. Barone$^{1}$, B. L. Altshuler$^{4}$ and F. Tafuri$^{1,2}$\\}

\affiliation{$^{1}$ Dip. Scienze Fisiche, Universit\`{a} di Napoli Federico II, Italy and {\sl Coherentia INFM-CNR}}
\affiliation{$^{2}$Dip. Ingegneria dell'Informazione, Seconda Universit\`{a} di Napoli, Aversa (CE), Italy}
\affiliation{$^{3}$ Dept. Microelectronics and Nanoscience, MINA, Chalmers University of Technology, S-41296 Goteborg, Sweden}
\affiliation{$^{4}$ Physics Dept.
Columbia University, New York NY 10027;\\ NEC Laboratories America INC, 4 Independence Day, Princeton, NJ 08554,USA}

\date{\today}

\begin{abstract}
Magneto-fluctuations of  the  normal  state resistance  $R_N$  have  been  reproducibly  observed   in high critical temperature 
superconductor ($HTS $) grain boundary junctions, at  low  temperatures. We  attribute  them  to 
 mesoscopic transport  in narrow channels across  the  grain boundary line. 
The  Thouless energy  appears  to be  the  relevant  energy  scale.   Our  findings have  significant implications
on quasiparticle  relaxation  and  coherent   transport in  $HTS$   grain  boundaries. 
\end{abstract}

\maketitle

 Junctions are extremely useful to test  very important properties  of
 high critical temperature superconductors ($HTS$)
 \cite{kirtley,hans,rop}, whose nature has not yet been completely established
\cite{natureph}.Recently high quality YBa$_{2}$Cu$_{3}$O$_{7-\delta }$
 ($YBCO$)
grain boundary ($GB$) Josephson junctions ($JJ$s) have given the first evidence of potentials to study exciting
crucial issues such as macroscopic quantum behaviors, coherence and dissipation \cite{mqt,nuovo}. The quantum tunnelling of such macroscopic $d-wave$ devices out of zero voltage state has been demonstrated along with the  quantization of its energy states. The cross-over temperature between quantum and classical regimes was found to  be of the order of $40 \: mK $.
This is evidence  of  the  fact  that  despite the short coherence lengths  in  a highly disordered  environment  and the  presence of low energy  (nodal) quasiparticles ( $qp$s)  due to the nodes of the $d-wave$ order parameter  symmetry \cite{kirtley,mqt,rop}, dissipation  in  a  $HTS$ junction  does not seem to be as disruptive for the quantum  coherence at  low temperatures, as one would naively expect. Increasing  the  available  information about  the   nature  of  $qp$s and their relaxation  processes is  of  crucial  importance  to unveil  the  mechanism  which leads to  superconductivity  in  $HTS$.

In  this  letter  we report on  transport  measurements of $YBCO$ biepitaxial ($BP$) $GB$ junctions (see scheme in Fig. $1a$), which  give  evidence of conductance fluctuations ($CF$s )  in  the magneto-conductance response of $HTS$ junctions. To our knowledge,  this is  the first time that $CF$s are measured in $HTS$ junctions. Our results prove the appearance of  mesoscopic  effects in the unusual energy regime  $k_BT << \epsilon _c < eV <\Delta$. 
Here $k_BT$  is the thermal  energy at  temperature $T$, $ \epsilon _c $ is  the
Thouless energy introduced below,  $V$  is  the  applied  voltage 
and $\Delta$ is the superconducting gap ($ \approx 20 \: meV $ for YBCO \cite{kirtley}). 

The progress registered in mesoscopic physics in the last two decades is impressive\cite{lee,aronov,today,imry}.  At low temperatures, quantum coherence can be monitored in  the conductance $G$ of a normal metallic sample of length $L_x$  attached  to  two reservoirs. The electron wave packets that carry current in a diffusive wire have minimum size of the order of $ L_T  > L_x >> l $ . Here $l$ is  the  electron mean  free  path in  the  wire  and  $L_T $ is  the  thermal  diffusion  length  $\sqrt{\hbar D/k_B T}$ ($D$  is  the  diffusion  constant).   The first inequality is  satisfied at relatively low temperatures as far as $k_B T << \epsilon_c \equiv  \hbar D /L_x^2 $. At low voltages ($ eV<< \epsilon _c $), the system is in the regime of universal conductance  fluctuations: the variance  $<\delta g^2 >$  of  the  dimensionless  conductance $g = G / 2e^2 /\hbar$  is of order of unity.   
Mesoscopic phenomena have been widely investigated  even  in metallic wires interrupted by tunnel junctions \cite{nazarov} and in Josephson junctions\cite{takay1,klapwijk,dubos}.
In experiments  on $Nb/Cu/Nb$  long junctions \cite{dubos},  phase coherence mediated by  Andreev reflection at the $ S/N$ interfaces, has been shown to persist  in  the whole
diffusive  metallic  channel, several hundreds of $nm$ long.
Such an effect is robust with respect to energy broadening due  to temperature, but it is 
expected to be fragile to energy relaxation  processes induced  by  the applied voltage. By contrast, 
in  our case, the $qp$ phase coherence  time  $\tau _\varphi$  seems  not  to  be limited by energy relaxation due to voltage induced  non-equilibrium. Mesoscopic  resistance oscillations are found even for $eV > \epsilon _c$, indicating that $qp$s do not loose coherence at low  temperatures, while diffusing across the $N$ bridge.
By  plotting  the  auto-correlation function  of  the measured $CF$ in magnetic  field $H$ (see  below), we  determine
$\epsilon _c $,  in analogy to similar studies of normal metal and semiconductor systems \cite{imry,quantumhall}. It is valuable that $HTS$ $GB$ $JJ$s make accessible this regime, which is not commonly achieved by conventional metals and junctions, providing a ``direct'' measurement of the Thouless energy.


\begin{widetext}

\begin{figure}[!htp]
    \centering
        \includegraphics[width=0.45\linewidth]{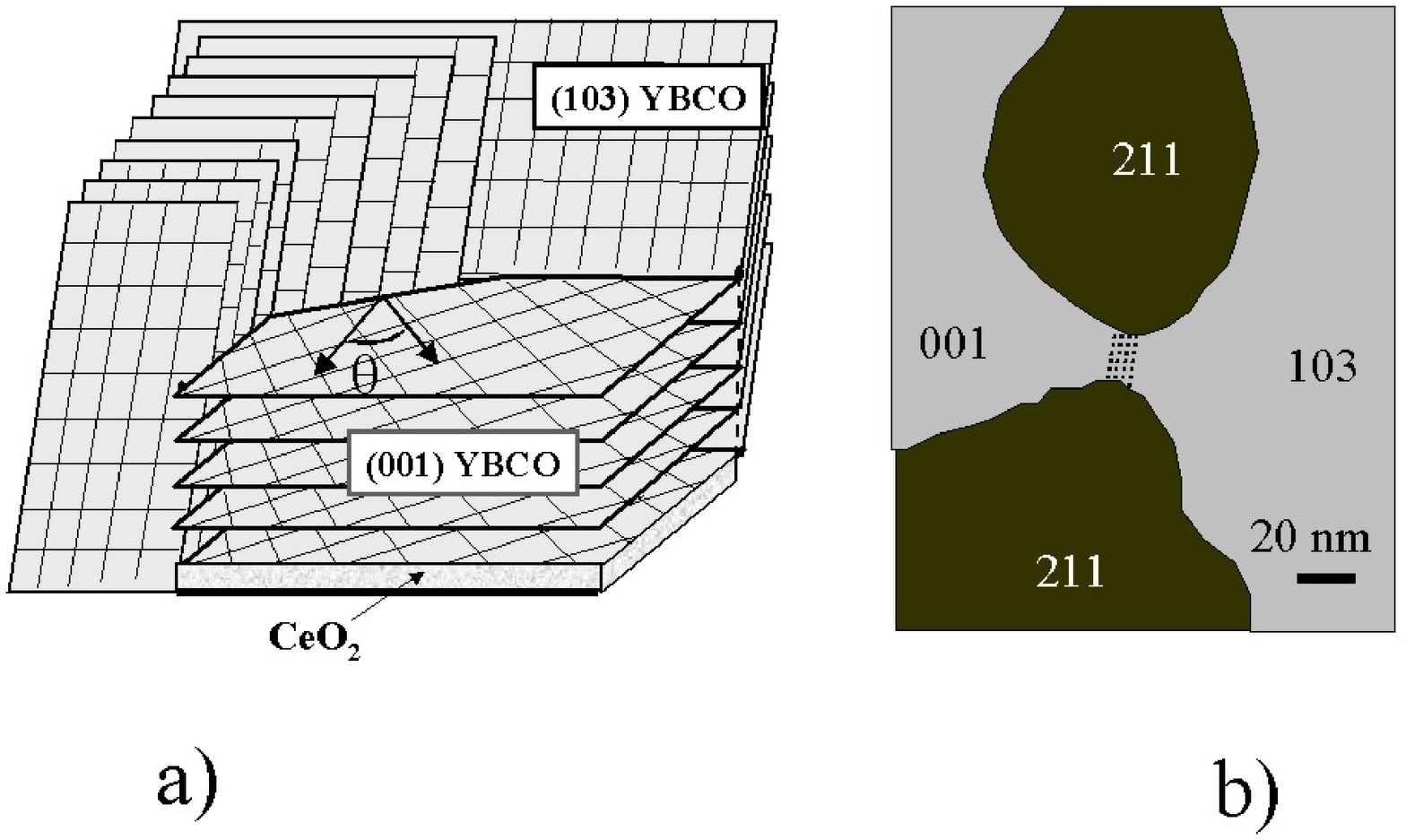}
        \includegraphics[width=0.45\linewidth]{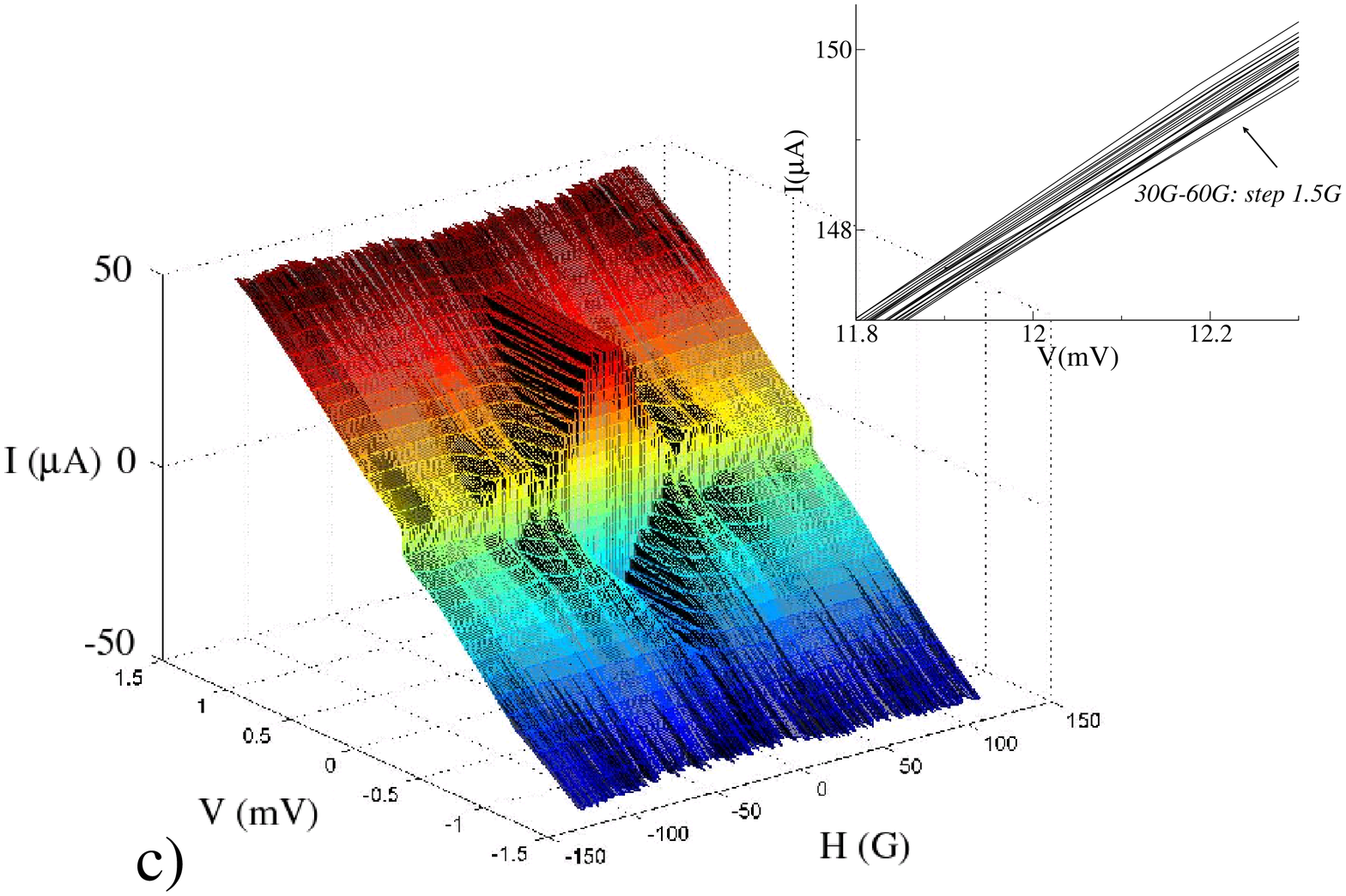}
\caption{(a) Sketch of the grain boundary structure for three different interface orientations.
(b) Sketch of a typical mesoscopic channel, that might be generated at the grain boundary line (dotted lines) because
of the presence of impurities. In biepitaxial junctions, the impurities may be due to Yttrium excess. (c) $ I-V $ curves as a function of the magnetic field $H$: the Fraunhofer pattern
indicates a uniform interface with critical current density $J_C \sim  8 \times  10^4\: A/cm ^2$ 
at $ T = 4.2 \: K $. In the inset, a few $ I-V $ curves are displayed at different magnetic fields
(from 30 G to 60 G, step 1.5 G) at high voltages to give a better evidence of the fluctuations of the resistance.}	
\label{fig1p}
\end{figure}
\end{widetext}

\begin{figure}[htb]	
\begin{center}		
\includegraphics[width=6cm]{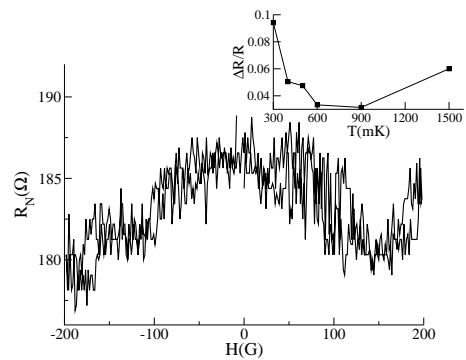}	
\end{center}	\caption{Resistance   $R_N $ $vs \: H $  at  $T = 300 \: mK $. Marked non  periodic  fluctuations are evident. In the inset, the zero field amplitude of the fluctuations
is reported  $vs \: T $.}
\label{fig2p}
\end{figure}

The $BP$ $JJ$s are obtained at the interface between a 
(103) $YBCO$ film grown on a (110) SrTiO$_3$ 
substrate and a $c$-axis film deposited on a (110) CeO$_2$  seed layer (Fig. $1a$). The 
presence of the CeO$_2$ produces an additional 45$^\circ$ in-plane rotation of the $YBCO$ axes 
with respect to the in-plane directions of the substrate. By suitable patterning of the CeO$_2$ seed layer, the interface 
orientation can be varied and tuned to some appropriate transport regime (in Fig. $1a$ we have indicated the two limiting cases
of 0$^\circ$ and 90$^\circ$, and an intermediate situation defined by the angle $\theta$, in our case $\theta = 60^\circ$) 
\cite{nuovo,tem}. We have selected junctions with sub-micron channels, as eventually confirmed by Scanning Electron Microscope 
analyses, and measured their $ I-V $ curves,  as a function of $ H $ and $ T $.  We  have investigated various samples, but here 
we focus mostly on the junction characterized  by the Fraunhofer magnetic pattern shown in Fig. $1c$, where it is more likely 
that only one uniform  superconducting channel is active \cite{nota}. In Fig.$1c$ the $I/V$ curves are reported in a $3d-$plot as a function of magnetic field $H$. The critical current ($I_C$) oscillations point to a flux periodicity which is roughly consistent with the typical size ($10-100 \: nm $) that we expect for our microbridge from structural investigations ( see  Fig. $1b$). We have to take into account that the London penetration depth in the (103) oriented electrode is of the order of a few microns (much larger than the one in c-axis YBCO films)(see \cite{mqt} for instance).

In low-$T_c$ $JJ$s,  $R_N$ represents the resistance at  a voltage a few  times  larger   than  the gap  values. This matter is more delicate when dealing with $HTS$ $JJ$s, because of the deviations from the 
ideal RSJ-like behavior \cite{rop,hans}.
In our case a linear branch typical of the  RSJ 
behavior starts at $ V=  5 \: mV$. 
Values of $V$ between $ 10 \: mV $  and $ 15 \: mV$ are 
representative for the problem we are investigating (see the inset of Fig. $1c$). The average resistance  $\overline{R}_N $ over the whole magnetic field range is $\sim 182 \:\Omega $. 
We choose $V \approx 12 \: mV $ for deriving the $R_N$  $vs. \: H $ curve, that is reported in Fig. $2$ 
at the temperature $T= 300\: mK$ in the interval  $[-200 \: G,\:200 \: G]$. 
AC magneto-conductance 
measurements basically give the same results, but  provide a more reliable 
Fourier transform,  as  we  will discuss  elsewhere. The dynamical resistance  is measured at fixed bias current. 
Conductance fluctuations, as those  shown in Fig. $2$ become appreciable at temperatures below $900 \: mK $ $ (k_BT <<  \hbar I_C/2e  )$, in the 
whole magnetic field range. Their amplitude  is   $ \sqrt{\langle (\delta G )^2 \rangle }  \sim 0.07 \:
\overline{G}_N $ (with $\overline{G}_N \equiv (\overline{R}_N)^{-1} )$ at $300 \: mK $  and are more than one order
of magnitude larger  than the noise. 
Below    $1.2 \: K $, the fluctuations are nonperiodic, and for sure not
related to the $I_C (H)$ periodicity, and  have all the typical
characteristics of   mesoscopic  fluctuations. 
The pattern shows occasionally some hysteresis  mostly at high magnetic fields,  when  sweeping  $H$ back,
which we attribute to some delay in the magnetic response. Above  $1.2 \: K$, thermal fluctuations dominate (see the inset in  Fig. $2$).
%

By performing the  ensemble average of $G_N$ over $H$, additional insights 
can be gained.
The auto-correlation function is defined  as:
\bea
\Delta G(\delta V)\:=  \hspace{5cm}\nonumber\\= 
  \frac{1}{N_{V}} \sum _{V} \sqrt {\left \langle (G_{V +
 \delta V}(H)\:-\:\overline{G}_N)(G_V(H)\:-\:\overline{G} _N )\right 
\rangle }_H
\label{auto}
\enea
 with $\Delta G(0) =  \sqrt{\langle (\delta G )^2 \rangle}$.
Here $N_V$ is the number 
of $\delta V$ intervals in  which  the range  $V\in (10-15 \: mV) $ has  been  divided. $\Delta G(\delta V)$
is reported in Fig. $3$ at various temperatures.  The plots are the results  of  an  average  over  
various  series  of  measurements, to suppress sample specific effects.

\begin{figure}[!htb]	
\begin{center}		
\includegraphics[width=6cm]{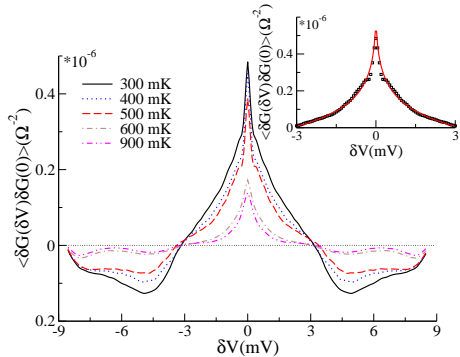}	
\end{center}	
\caption{The autocorrelation function of  Eq.(\ref{auto}) is derived from experimental data
for different temperatures. In the inset the autocorrelation function at 
T= 300 mK is fitted on the basis of Eq.\ref{fit} (full  line)}
\label{fig3p}
\end{figure}

As we show in  the inset  of  Fig. $3$, this  autocorrelation  function is
well  described by  a  fit  
based  on  the  standard theory for transport of diffusions modes in 
  mesoscopic  channels \cite{aronov}.
The  effect of  the  applied  $ H $ on  the interference  is 
expected  to rule  out  contributions  to  the  conductance  fluctuations 
due to cooperons.
The  autocorrelation as a function of the phase change 
$\Delta  \varphi$  is: 
\bwt
\begin{eqnarray} 
\left \langle 
\delta G( \Delta \varphi ) \delta G (0 )  \right \rangle =  {\sum _{m=-\infty } ^\infty} ' \langle G_m G_{-m}  \rangle \:
 e^{im\Delta\varphi } \approx  
\frac{2{\cal{N}}}{3 \pi}\:   \: \left ( \frac{2e^2}{h}\right )^2
 \frac{\epsilon_c}{k_B T} \: \frac{L_yL_z}{L_x L_T} \: \: \Re  e \: \left
 \{\ln  {\left ( 1-e^{- (\xi  +i\Delta \varphi )}
\right )}\right \}  \:\:  
\label{fit}
\enea
\ewt
Here the  parameter $\xi=\pi  L_x /L_T$ contains the important information on the channel effective length with respect to $L_T$ 
( ${\cal{N}}$ is a constant of the order 1,  up to higher orders  in  $ e^{-  |m|\xi }$).
In  deriving  eq.(\ref{fit}) we  have  assumed  $
k_B T >> \hbar /\tau_\varphi $ and larger than   
  the  level spacing in  the  microbridge \cite{spivak2}. 
 The Fourier  components of the auto-correlation  $ \langle G_m G_{- m}  \rangle $,  in  the
  limit  of  small  $m$'s, are  
 $\langle G_m G_{m'}  \rangle   \propto \: \delta _{m',-m} \:   erfc\: \left ( \sqrt{  |m| \xi }\right )  
 \:  e^{-  |m|\xi } / |m|    $.
The  exponential  integral  function $erfc $,  together  with 
  the $\delta _{m',-m} $, arises  from  the  sum  
over  the  diffusion  modes.
 Correlations  at  larger $\Delta \varphi$  have  to blurr out  due to other 
mechanisms of  dephasing  not  included  here  and cannot  follow  eq.(\ref{fit}).
 In the  inset  of Fig. 3 the full line shows the 
best fit obtained  by using Eq.(\ref{fit}) with 
 $ \Delta \varphi   = \sqrt{e \delta V /\epsilon_c }$ and 
$\xi  = 0.1$.  
 This value of $\xi  = 0.1$ is consistent with a $L_x$ value of the order of 50 nm.
The  value of $\epsilon_c$ can be  read out of 
the plot from the location of the node, giving $\epsilon_c $ of  the  order 
of  $ 3 \: meV $. 
Our  plots  can  be compared  to  the ones  of  ref.\cite{nazarov} obtained for  a  $Al/Al_2O_3/Al $ tunnel  junction  at   $ T=20 mK $, $V=0.8 mV $  in  a magnetic field  $H=0.5 Tesla $ which  drives  the  junction  into  the  normal  state.  In ref.\cite{nazarov},  the energy broadening extracted from the experiment, $ \sim max\{ \hbar /\tau _\varphi , \epsilon _c \} $,  is $2.3 \: \mu  eV $, at  $20  mK $. 

 In  our  case,   $ < \left  (\overline{\Delta  \varphi} \right )^2 >
\approx e \overline{\delta V} / \epsilon _c \sim  1  $  fixes  the  scale of amplitude  of the 
voltage fluctuations  $\overline{\delta V }$  for  carriers  diffusing  across the bridge.   It is remarkable that the coherence time which is expected when dephasing is dominated by  small-energy-transfer (Nyquist) 
  scattering of nodal $qp$s ( $\sim   \hbar /V$ \cite{ingold}) is  much shorter than  $\hbar /\epsilon _c $. The voltage  drop
appears to  occur  mainly away from  the microbridge  as the  electrochemical 
 potential cannot  change  significantly   over  distances  smaller than  the  
wave packet  size \cite{zhou,ludwig}.
We  have  found   that both  temperature and   bias voltages up to  $30 \: meV $ reduce the  amplitude  of  the  
$CF$s ( see the  inset of  Fig. 2)  and of $\epsilon _c $,  but they do not wash out the correlations.  $ \sqrt{\left \langle \left (\delta G /\overline{G}_N
 \right )^2\right \rangle }$ drops  with  temperature, according to eq.(2)  as  probed from  the  experiment (see inset of  Fig.2).

Given that the thermal diffusion length  $ L_T  \sim 0.13 \: \mu m $ at $ 1\: K $ (with $D\sim 20\div  24 \: cm^2/sec $  for $YBCO$\cite{gedik}),  the condition
$L_x < L_T $ is certainly  met, thus confirming that $\epsilon _c $ is the relevant energy scale. The phase  coherence length  exceeds  the size of the  microbridge  and therefore thermal decoherence can be ruled out.  Hence,  not  only  the   interference  of  tunnelling currents takes  place, which provides  the   Fraunhofer  pattern of  Fig. 1 $c$,  but  also  the 
quantum  interference  of  electrons    returning  back  to  the  junction in  their diffusive  motion at the $GB$.   
We  believe, this  is  the  first  time  that  the wave-like   nature of  the  carriers  fully appears  in $HTS$  systems  and  that  mesoscopic  scales  can be identified.

As a final remark, our measurements show that $\epsilon_c$ is the relevant energy scale for the supercurrent  as well. Indeed, we find that $e I_c R_N $ and   $\epsilon_c$ are of  the  same  order  of  magnitude, in agreement with the typical values measured in $HTS$ $JJ$s \cite{hans,rop} and  the results on diffusive phase-coherent normal-metal $SNS$ weak links \cite{dubos,wilhelm}. This  feature gives additional  consistency to the  whole  picture, relating  the critical  current, which  is mediated by  subgap Andreev reflection, with the transport properties at high voltages. The  coherent diffusive regime across the $S/N/S$ channel of our GB junctions   persists  even  at  larger  voltage bias  $  \Delta  > eV > \epsilon _c$ and   is the  dominating  conduction  mechanism. 
Hence, microscopic features of the  weak link appear as less relevant, in favour of mesoscopic, 
non local  properties. The results define important attributes of the role of grain boundaries in the 
transport in $HTS$ junctions and in particular of the narrow
self-protected channels formed in the complex growth of the oxide grain  boundary \cite{gross,buhrman,hans,rop}. 
The fact that  the dominant energy  scale  is found to be $\epsilon _c \sim 3 \: meV $ sets  a lower bound to the  relaxation time at low $T$ for  $qp$s of energy  even ten  times  larger, of the order of  picoseconds. It has been argued that antinodal $qp$s may require slow diffusive drift of momentum along the Fermi surface towards  the nodes of the $ d-wave $ gap \cite{gedik,howell}.

In  conclusion,  we  have given  evidence  of  conductance  fluctuations in
 $HTS $ grain  boundary  Josephson  Junctions constricted   by  one 
single  micro-bridge   at  low  temperatures.   $CF$s  are  the  signature of a coherent diffusive regime. Our  results  seem to suggest an unexpectedly long $qp$s phase coherence time and represents another strong indication that the role of dissipation in HTS has to be revised.

\vspace*{0.5cm}

Work partially supported by the ESF projects "$\Pi$-Shift" and "QUACS" and by MIUR funds (Italy). Numerical  help by 
B.Jouault  and P.Lucignano, as  well  as  discussions  with T.Bauch,  C.Biagini, G.Campagnano, J.R. Kirtley, A.Mirlin, H.Poithier and Y. Nazarov  are gratefully  acknowledged. 



\end{document}